\documentclass[conference, a4paper]{IEEEtran}
\ifCLASSINFOpdf
  \usepackage[pdftex]{graphicx}
  \graphicspath{{../pdf/}{../jpeg/}}
  \DeclareGraphicsExtensions{.pdf,.jpeg,.png}
\else
\fi
\usepackage[cmex10]{amsmath}
\usepackage{amsthm}
\usepackage{amssymb}
\theoremstyle{definition}

\theoremstyle{remark}

\ifCLASSOPTIONcompsoc
  \usepackage[caption=false,font=normalsize,labelfont=sf,textfont=sf]{subfig}
\else
  \usepackage[caption=false,font=footnotesize]{subfig}
\fi
\usepackage{graphicx}

\usepackage{bm}

\usepackage{graphicx}

\usepackage{cite}

\begin{document}
\bstctlcite{IEEEexample:BSTcontrol}

%
\title{An Alternative Estimation of Market Volatility\\
based on Fuzzy Transform}

\author{\and
\IEEEauthorblockN{Luigi Troiano}
\IEEEauthorblockA{
University of Sannio\\
Department of Engineering \\
Benevento, Italy\\
Email: troiano@unisannio.it}
\and
\IEEEauthorblockN{Elena Mejuto}
\IEEEauthorblockA{
University of Sannio\\
Department of Engineering \\
Benevento, Italy\\
Email: mejutovilla@unisannio.it}
\and
\IEEEauthorblockN{Pravesh Kriplani}
\IEEEauthorblockA{
University of Sannio\\
CISELab\\
Benevento, Italy\\
Email: pravesh.kriplani@ciselab.org}
}


%


\maketitle

\begin{abstract}
Realization of uncertainty of prices is captured by volatility, that is the tendency of prices to vary along a period of time. This is generally measured as standard deviation of daily returns. In this paper we propose and investigate the application of fuzzy transform and its inverse as an alternative measure of volatility. The measure obtained is compatible with the definition of risk measure given by Luce. A comparison with standard definition is performed by considering the NIFTY 50 stock market index within the period Sept. 2000 - Feb. 2017.

\end{abstract}


%
\IEEEpeerreviewmaketitle

\section{Introduction}

The value of financial instruments is subject to change over time. Because of that, risk can result into possible losses due to price fluctuations. Risk associated to uncertainty of prices is a feature to keep under control in almost all activities, including trading strategies, portfolio management, and asset allocation. Although uncertainty is driven by factors external to markets, such as investor sentiment, economic fundamentals, macroeconomic environment, future expectations, etc., the way it affect the dynamics of prices represents the source of risk we are considering in this context.

Among the different ways to look at risk, volatility of asset prices is widely accepted as a realization of the uncertainty in financial markets. Volatility refers to the degree of price fluctuations. Volatility is not directly observable. It is a tendency of price to move and different definitions have been proposed in order to measure it. In the context of this paper we will refer to \emph{historical} volatility, that is the realized volatility of a financial instrument over a given time period, against \emph{implied} volatility, that is an estimation of volatility in the future looking at the price of options.

A common measure of historical volatility is defined as standard deviation of returns, that is obtained by variance
\begin{equation}
\vartheta = E[|r-\mu|^2]
\end{equation}
where $r$ represents the returns over a given period of time $T$ and $\mu$ their mean in the same period. This is a specific case of other common ways to measure volatility that can be generalized as
\begin{equation}
\vartheta_\theta = E[|r-\mu|^\theta]
\end{equation}
When $\theta = 1$, volatility is measured as mean absolute deviation. Returns can be absolute, i.e., $r = p_{new} - p_{old}$. But more generally they are relative, i.e., $r = (p_{new} - p_{old})/p_{old}$. In practice, it is also common to refer to logarithmic returns, that are computed as $r = log(p_{new}/p_{old})$.

\IEEEpubidadjcol

Grenger \cite{JAE:JAE687} argues that volatility measures given above are related to the risk measure developed by Luce \cite{Luce1981}. In this model, a measure of risk $R$ is given as
\begin{equation}\label{eq:grenger}
R(f) = K_1 \int\limits_0^{+\infty} |\rho|^\theta f(\rho)d\rho + K_2 \int\limits^0_{-\infty} |\rho|^\theta f(\rho)d\rho
\end{equation}
where $K_1$ and $K_2$ are two constants, both $\geq 0$, and $\rho$ is referred to returns, whose distribution is given by a density function $f$. This definition satisfies two important axiomatic properties given by Luce \cite{Luce1980}. First, it is proportional to any change of scale of returns, meaning that if we consider $\rho' = \alpha \rho$ (with $\alpha > 0$), we will get a distribution of returns given by 
\begin{equation}
f_\alpha(\rho') = \frac{1}{\alpha} f( \frac{\rho'}{\alpha} ) = \frac{1}{\alpha} f( \rho )
\end{equation}
Therefore, the associated risk is
\begin{equation}\label{eq:luce1}
R(f_\alpha) = \Gamma(\alpha) R(f)
\end{equation}
where $\Gamma(\alpha)$ is a monotonic increasing function, assuming $\Gamma(1) = 1$ for consistency. Second, there is a non negative function $U$, with $U(0) = 0$, such that
\begin{equation}\label{eq:luce2}
R(f) = \int\limits_{-\infty}^{+\infty} U(f(\rho)) d\rho
\end{equation}
where $U$ is expression of punctual measure of risk associated to the probability of having $\rho$ as return. Eq.\eqref{eq:luce2} becomes Eq.\eqref{eq:grenger} if $K_1 = K_2 = K$ and $U(f(\rho)) = K|\rho|^\theta f(\rho)$. 

In the attempt of providing a measure of volatility different from the common approach based on the standard deviation, we investigate the application of fuzzy transform. Fuzzy transform \cite{Perfilieva:2006} is gaining interest for its characteristics in terms of universal approximation, smoothing and reconstruction \cite{Loia2017209,Tomasiello2015,Troiano201211,Troiano2011121,Troiano2010379}. We aim to exploit these features for volatility estimation. This contribution is organized as follows: Section II provides some preliminaries regarding fuzzy transform; Section III presents the model of volatility measurement based on fuzzy transform; Section IV discusses experimental results; Section V outlines conclusions and future directions. 

\section{Preliminaries}

Let $I=[a,b]$ be a closed interval and $x_1,x_2,\ldots ,x_n$, with $n\ge 2$, be points of $I$, called nodes, such that $a=x_1<x_2< \ldots <x_n=b$. We consider as partition of $I$, a collection of ${A_1, A_2,\ldots , A_n }$ of fuzzy sets, with $A_i :I \rightarrow [0, 1]$, with $i = 1, \ldots , n$ such that
\begin{itemize}
\item $A_i(x)=0$ if  $x \in [a, x_{i-1}]\cup [x_{i+1},b]$
\item $A_i(x) > 0$ if  $x \in (x_{i-1}, x_{i+1})$
\item $A_i(x_i) = 1$
\item $A_i(x) \leq A_i(x')$ for any $x,x' \in [x_{i-1}, x_{i}], x < x'$ 
\item $A_i(x) \geq A_i(x')$ for any $x,x' \in [x_{i}, x_{i+1}], x < x'$
\item $\sum\limits_{i=1}^n A_i (x) = 1$, for all $x \in I$
\end{itemize}

The fuzzy sets  ${A_1, A_2,\ldots , A_n }$ are called basic functions. If the nodes are equidistant, the partition is said \emph{uniform}. The fuzzy partition can be obtained by means of several basic functions. The most common are the hat functions
\begin{equation}
A_j(x)=\left\{\begin{array}{rr}
{(x_{j+1}-x)/ (x_{j+1}-x_j),}& {x \epsilon [x_j, x_{j+1}]}\\
{(x - x_{j-1})/(x_j - x_{j-1}),}& {x \epsilon [x_{j-1}, x_j]}\\
{0,}& {otherwise}
\end{array}\right.
\end{equation}
and the z-shaped basic functions
\begin{equation}
A_j(x)=\left\{\begin{array}{rr}
{{1\over 2} \left(\cos(\pi{{x-x_j}\over{x_{j+1}-x_j}})+1\right),}& {x \epsilon [x_j, x_{j+1}]}\\
{{1\over 2} \left(\cos(\pi{{x-x_j}\over{x_j-x_{j-1}}})+1\right),} & {x \epsilon [x_{j-1}, x_j]}\\
{0,} & {otherwise}
\end{array}\right.
\end{equation}

The fuzzy transform (F--transform) \cite{Perfilieva:2006} of a function $f(x)$ continuous on $I$ with respect to  $ \{A_1, A_2,\ldots , A_n \}$ is the $n$--tuple $[F_1,F_2,\ldots,F_n]$ whose components are computed as 
\begin{equation}
F_i={{\int_a^b f(x) A_i(x)dx}\over{\int_a^b A_i(x)dx}}.
\label{eq1}
\end{equation}
They minimize the following error functional
\begin{equation}
\Phi=\int_a^b(f(x)-F_i)^2 A_i(x)dx,
\label{eq_phi}
\end{equation}

The inverse F--transform of $f$ is defined as
\begin{equation}
f_{F,n}(x) =\sum_i^n F_i A_i(x),\qquad x\epsilon I
\label{eq2}
\end{equation}
It provides a smoothed approximation of function $f$.

In the case of time series, where the function $f$ is known only at a given set of points $\{t_1,t_2,\ldots,t_m\}$, we use the discrete F--transform, whose components are defined as 
\begin{equation}
F_i={{\sum\limits_{j=1}^m f(t_j) A_i(t_j)}\over{\sum\limits_{j=1}^m A_i(t_j)}}, \qquad i=1..n
\label{eqfd}
\end{equation}

The discrete inverse F--transform is defined as
\begin{equation}
f_{F,n}(t_j)=\sum_i^n F_i A_i(t_j),\qquad j=1..m
\label{eqfi}
\end{equation}

\section{Measuring the risk}

Let us assume daily returns expressed as
\begin{equation}
r_t = \frac{p_t - p_{t-1}}{p_{t-1}}
\end{equation}
where $p_t$ is the price at day $t$ and $p_{t-1}$ the price the day before. Logarithmic returns can be expressed as
\begin{equation}
r_t = \log{\frac{p_t}{p_{t-1}}}
\end{equation}

Given a time horizon $T$, historical volatility is computed as
\begin{equation}
\sigma = \sqrt{E[R_T^2]- E[R_T]^2}
\end{equation}
that is the standard deviation of daily returns. The random variable $R_T$ is represented by a sample of returns $r_t$ with $t \in T$. Depending on the extension of $T$, we have different measures of volatility. In general, volatility is computed by looking back at past year, past quarter, past month or past week.\footnote{Standard deviation is computed over the daily returns. In order to compare volatility when computed over a different time resolution, we have to refer to a time frame in common. Generally it is used the year, so that $\sigma_A = \sigma \sqrt{T_A}$ where $T_A$ is the number of times the period falls in a year. This process is called annualization. If standard deviation is computed according to the daily returns, $\sqrt{T_A} = \sqrt{252}$ that is the number of trading days in a year. We note that $T_A$ is related to the frequency by which returns are collected and not to the extent $T$ by which we look at past values.}

We consider the application of fuzzy transform, following the procedure that is outlined below.

\subsection{Partition}

It is possible to choose any convenient uniform fuzzy partition of time. As general rule, the support of sets should be such that cardinality is preserved, that is
\begin{equation}\label{eq:card_cond}
\sum\limits_{j=1}^m A_i(t_j) = \sum\limits_{t \in supp(A_i)} A_i(t) = T \qquad i = 1 .. m
\end{equation}

For the sake of simplicity, we consider triangular fuzzy sets $A_i(t)$. This means that nodes should be distributed uniformly at distance $T$ each other, in order to meet the condition expressed by Eq.\eqref{eq:card_cond}, or equivalently that $supp(A_i) = [\tau_i - T, \tau_i + T ]$ where $\tau_i$ is the time node associated to the fuzzy set $A_i$.

\subsection{Baseline}

We compute the fuzzy transform of daily returns as
\begin{equation}
B_i = \frac{1}{T} \sum\limits_{t = \tau_i - T}^{\tau_i + T} r_t A_i(t)
\end{equation}

The baseline is computed by means of the inverse transform as
\begin{equation}
b_t = \sum\limits_{i = 1}^{n} B_i A_i(t)
\end{equation}

The baseline plays the same role as the mean in computing the standard deviation.

\subsection{Deviation}

First we compute the fuzzy transform of absolute returns as
\begin{equation}
H_i = \frac{1}{T} \sum\limits_{t = \tau_i - T}^{\tau_i + T} |r_t| A_i(t)
\end{equation}
and its inverse
\begin{equation}
h_t = \sum\limits_{i = 1}^{n} H_i A_i(t)
\end{equation}

Finally, we assume as measure of volatility
\begin{equation}
d_t = h_t - b_t
\end{equation}

Making explicit the contribution provided by both terms, we get
\begin{equation}\label{eq:dev_explicit}
\begin{split}
d_t &= \sum\limits_{i = 1}^{n} \left( H_i - B_i \right) A_i(t) =
\frac{1}{T} \sum\limits_{i = 1}^{n} \left( \sum\limits_{t = \tau_i - T}^{\tau_i + T} |r_t| - r_t \right) A_i^2(t)\\
    &= \frac{1}{T} \sum\limits_{t = \tau_i - T}^{\tau_i + T} \left( (|r_t| - r_t) \sum\limits_{i = 1}^{n} A_i^2(t) \right) 
    \leq \frac{1}{T} \sum\limits_{t = \tau_i - T}^{\tau_i + T} (|r_t| - r_t)
\end{split}
\end{equation}
being $A_i^2(t) \leq A_i(t)$ as $A_i(t) \in [0,1]$, thus
\begin{equation}
0 \leq \Lambda^2 = \sum\limits_{i = 1}^{n} A_i^2(t) \leq \sum\limits_{i = 1}^{n} A_i(t) = 1 \qquad t \in \{t_1,\ldots,t_m\}
\end{equation}

\subsection{Properties of deviation}

It is trivial that $d_t \geq 0$. But it is more interesting to observe that it fulfills both conditions laid down by Luce \cite{Luce1980,Luce1981} to be an appropriate measure of risk. Indeed, with respect to the first condition (see Eq.\eqref{eq:luce1}), if $r'_t = \alpha r_t$ (with $\alpha > 0$), we get
\begin{equation}
B'_i = \frac{1}{T} \sum\limits_{t = \tau_i - T}^{\tau_i + T} r'_t A_i(t) = 
\frac{1}{T} \sum\limits_{t = \tau_i - T}^{\tau_i + T} \alpha r_t A_i(t) = \alpha B_i
\end{equation}
and
\begin{equation}
b'_t = \sum\limits_{i = 1}^{n} B'_i A_i(t) = \sum\limits_{i = 1}^{n} \alpha B_i A_i(t) = \alpha b_t
\end{equation}
In addition,
\begin{equation}
H'_i = \frac{1}{T} \sum\limits_{t = \tau_i - T}^{\tau_i + T} |\alpha r_t| A_i(t) = \alpha H'_i
\end{equation}
and its inverse
\begin{equation}
h'_t = \sum\limits_{i = 1}^{n} \alpha H_i A_i(t) = \alpha h_t
\end{equation}

As a consequence, we obtains
\begin{equation}
d'_t = h'_t - b'_t = \alpha( h_t - b_t ) = \alpha d_t
\end{equation}

With respect to the second condition (see Eq.\eqref{eq:luce2}), we consider the set $\textbf{R} = \{ r_t \} $, with $t \in [\tau_i-T, \tau_i+T]$. Since the time series is discrete, $\textbf{R}$ is finite. We can rewrite Eq.\eqref{eq:dev_explicit} as 
\begin{equation}
d_t = \frac{1}{T} \sum\limits_{t = \tau_i - T}^{\tau_i + T} (|r_t| - r_t) \Lambda^2
     = \frac{1}{T} \sum\limits_{r_t \in \textbf{R}} (|r_t| - r_t) \Lambda^2 \psi(r_t)
\end{equation}
where $\psi(r_t)$ is the number of times $r_t$ occurs in $\textbf{R}$. Since, data points are finite, also possible values of occurrences are finite, therefore
\begin{equation}
d_t = \sum\limits_{r_t < 0} \frac{1}{T} 2|r_t| \Lambda^2 \psi(r_t) + \sum\limits_{r_t \geq 0} \frac{1}{T} (r_t - r_t) \Lambda^2 \psi(r_t)
\end{equation}
In summary,
\begin{equation}\label{eq:d_decomposed}
d_t = \sum\limits_{r_t < 0} \frac{2}{T} \Lambda^2 \psi(r_t) \cdot |r_t| = \sum\limits_{r_t < 0} \frac{2q}{T} \Lambda^2 \varphi(r_t) \cdot |r_t|
\end{equation}
where $\varphi(r_t) = \frac{\psi(r_t)}{q}$ and $q$ is the number of data points. Thus, $\varphi(r_t)$ is the frequency of return $r_t$. On the other side,
\begin{equation}\label{eq:luce2FT}
\begin{split}
\int\limits_{-\infty}^{+\infty} U(\varphi(r_t)) dr_t &= \sum\limits_{r_t \in \textbf{R}} U(\varphi(r_t)) r_t \\
&= \sum\limits_{r_t < 0} U(\varphi(r_t)) r_t + \sum\limits_{r_t \geq 0} U(\varphi(r_t)) r_t
\end{split}
\end{equation}
By comparing Eq.\eqref{eq:luce2FT} to Eq.\eqref{eq:d_decomposed}, we get
\begin{equation}
U(\varphi(r_t)) =
\begin{cases}
\frac{2q \Lambda^2}{T}\varphi(r_t) & r_t < 0 \\
0 & r_t \geq 0
\end{cases}
\end{equation}
This function is non negative and $U(0) = 0$ as required by Luce conditions. It is interesting to observe that the proposed measure of volatility is zero for positive returns. This is not the case of volatility measured as standard deviation. 

\begin{figure*}[t!]
\centering
\subfloat[Yearly]{ 
  \includegraphics[width=0.85\textwidth]{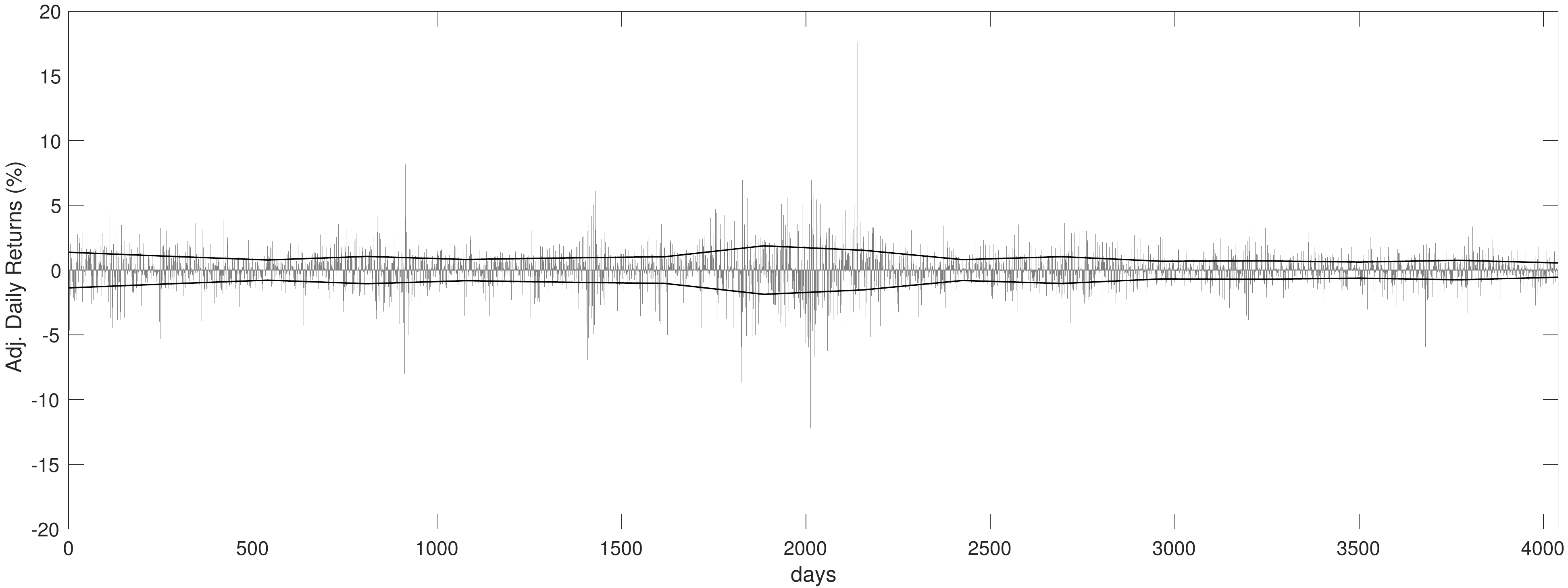}
}
\\
\subfloat[Monthly]{ 
  \includegraphics[width=0.85\textwidth]{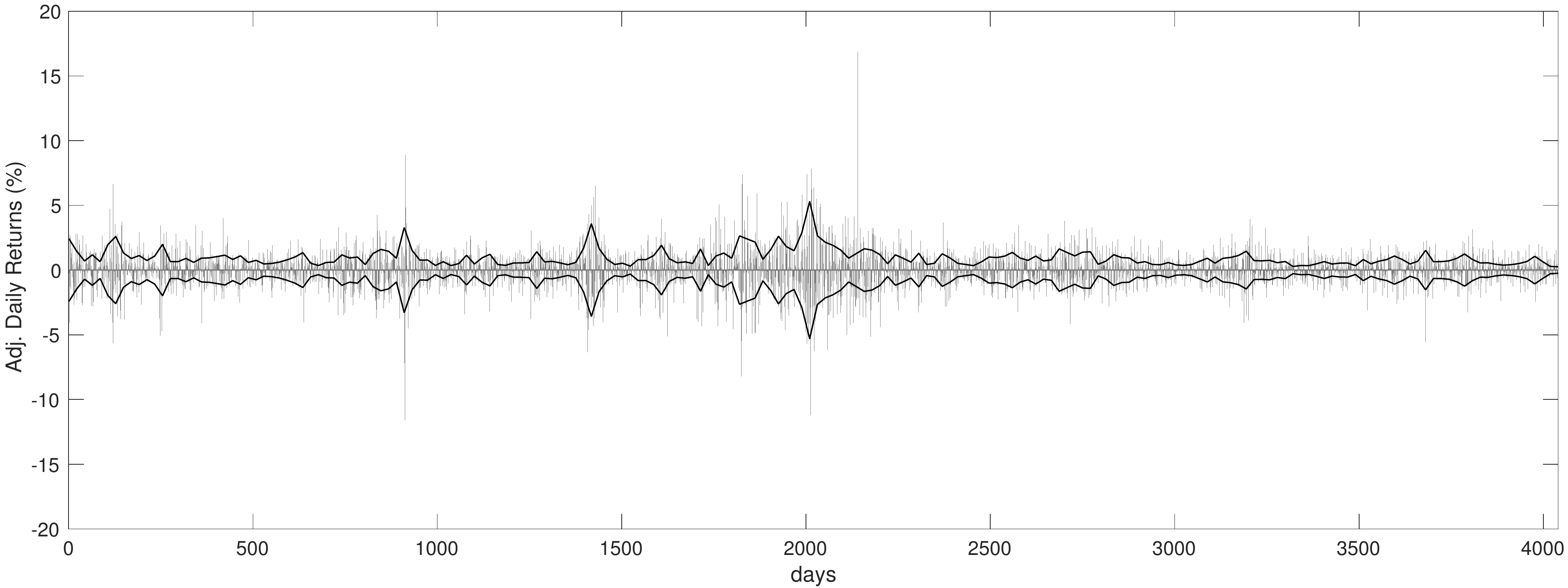}
}
\\
\subfloat[Weekly]{ 
  \includegraphics[width=0.85\textwidth]{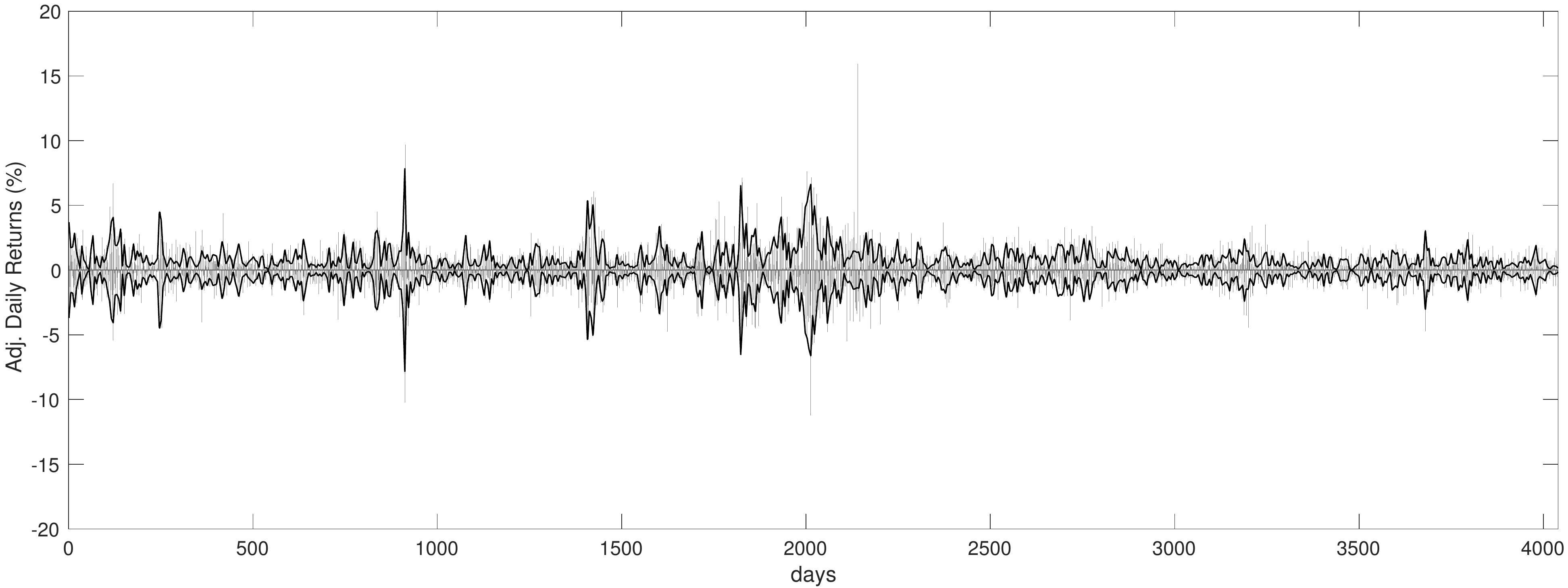}
}
\caption{Annual volatility computed by means of fuzzy transform}
\label{fig:FT_vol}
\end{figure*}

\begin{figure*}[t!]
\centering
\subfloat[Yearly]{ 
  \includegraphics[width=0.85\textwidth]{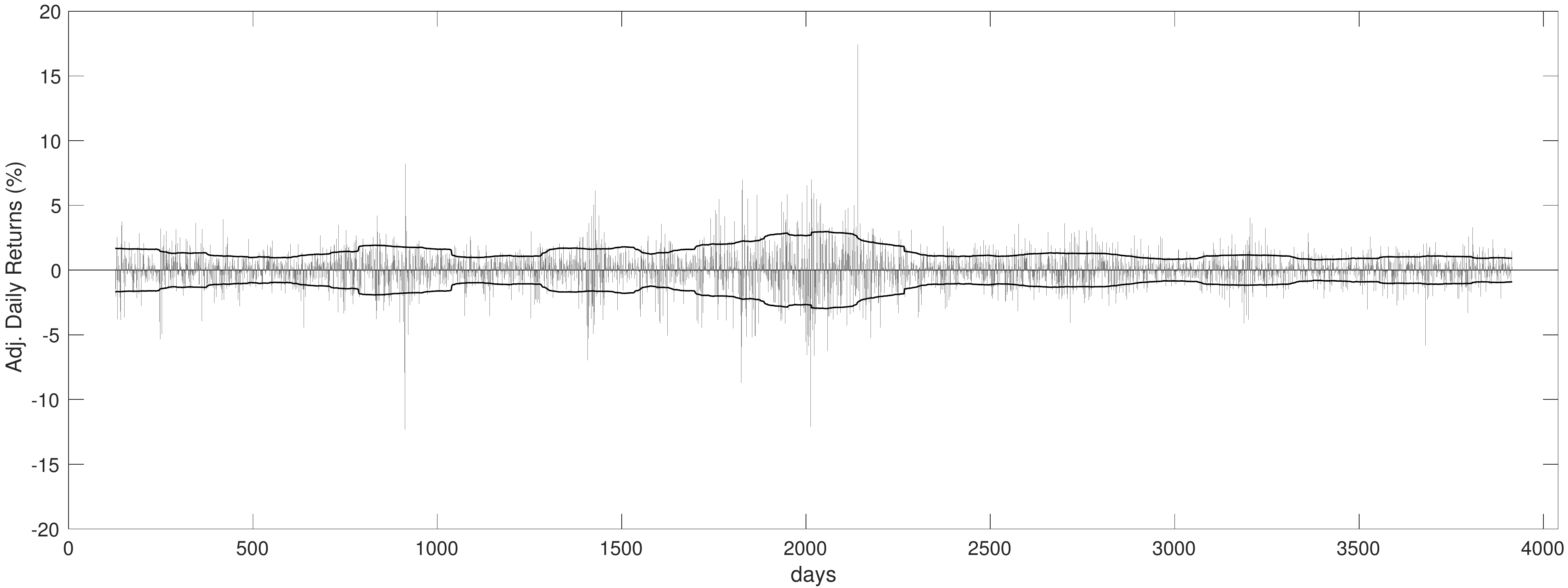}
}
\\
\subfloat[Monthly]{ 
  \includegraphics[width=0.85\textwidth]{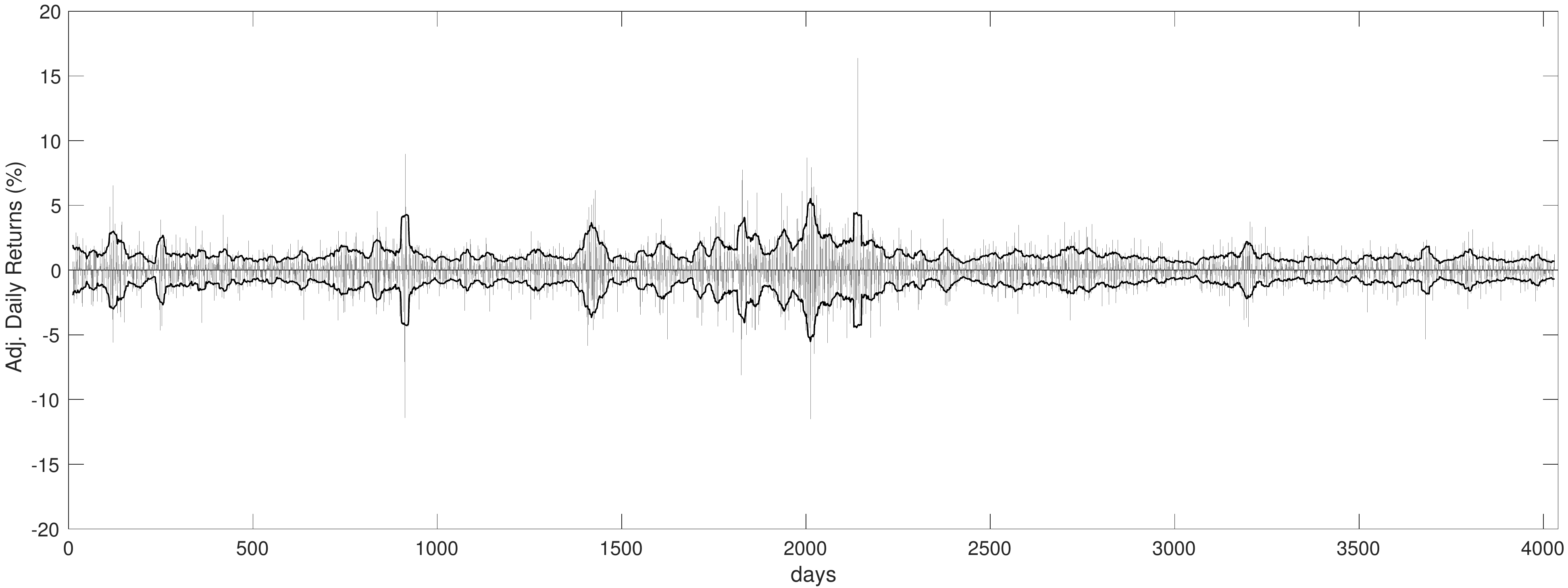}
}
\\
\subfloat[Weekly]{ 
  \includegraphics[width=0.85\textwidth]{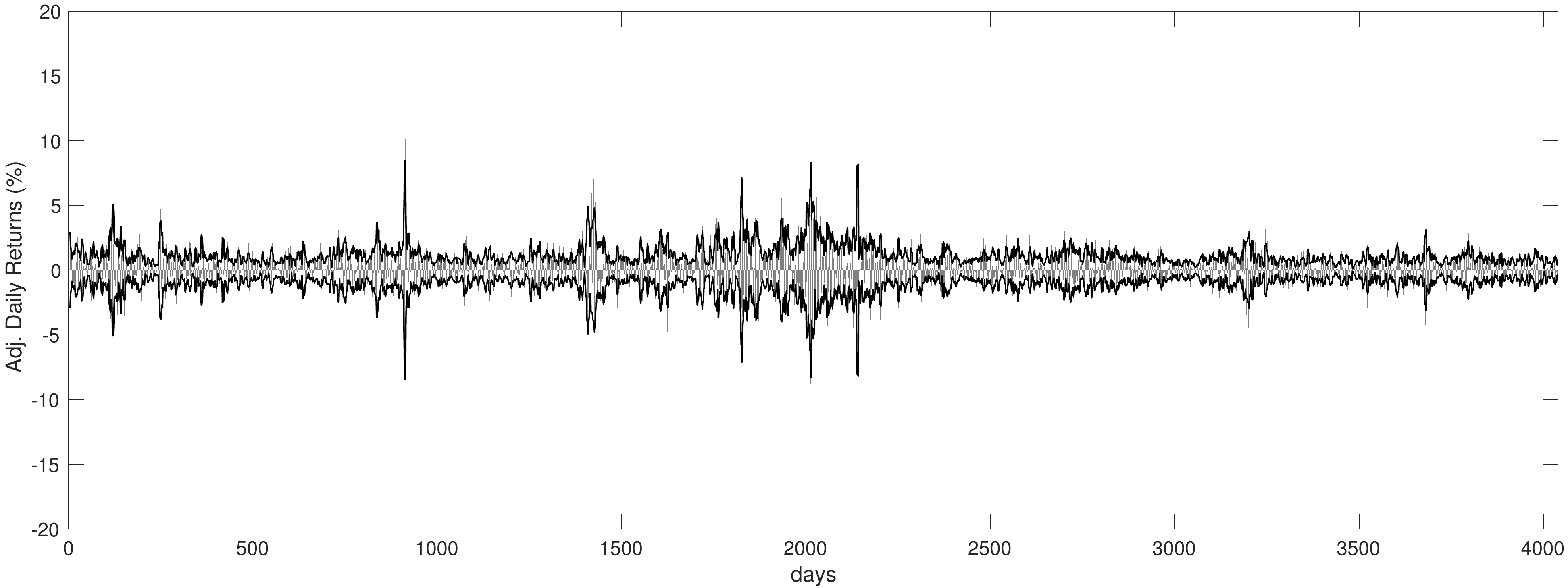}
}
\caption{Annual volatility computed by means of standard deviation}
\label{fig:STD_vol}
\end{figure*}

\begin{figure*}[t!]
\centering
\subfloat[Annual]{
  \includegraphics[width=0.6\columnwidth]{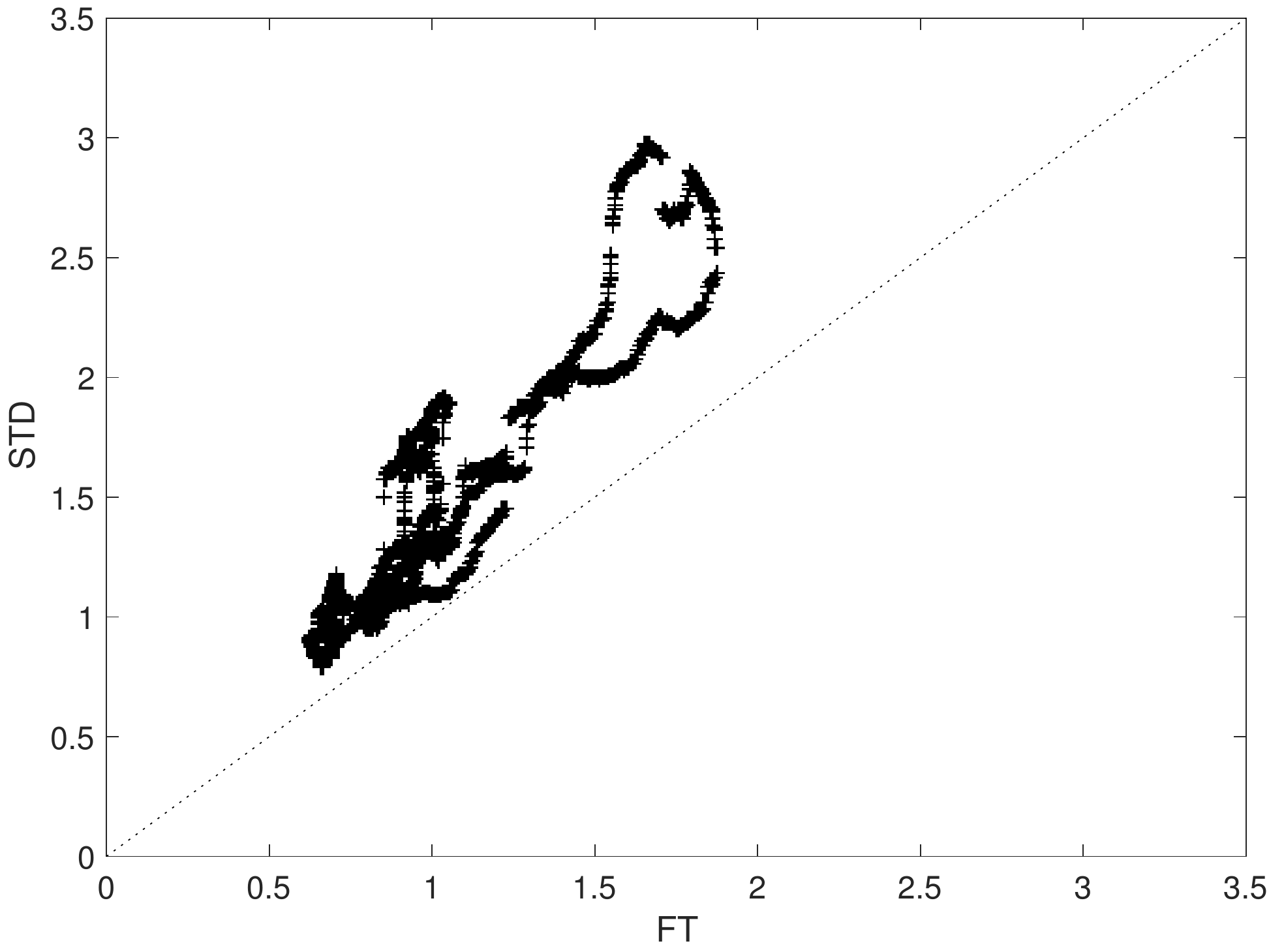}
}
\subfloat[Monthly]{
  \includegraphics[width=0.6\columnwidth]{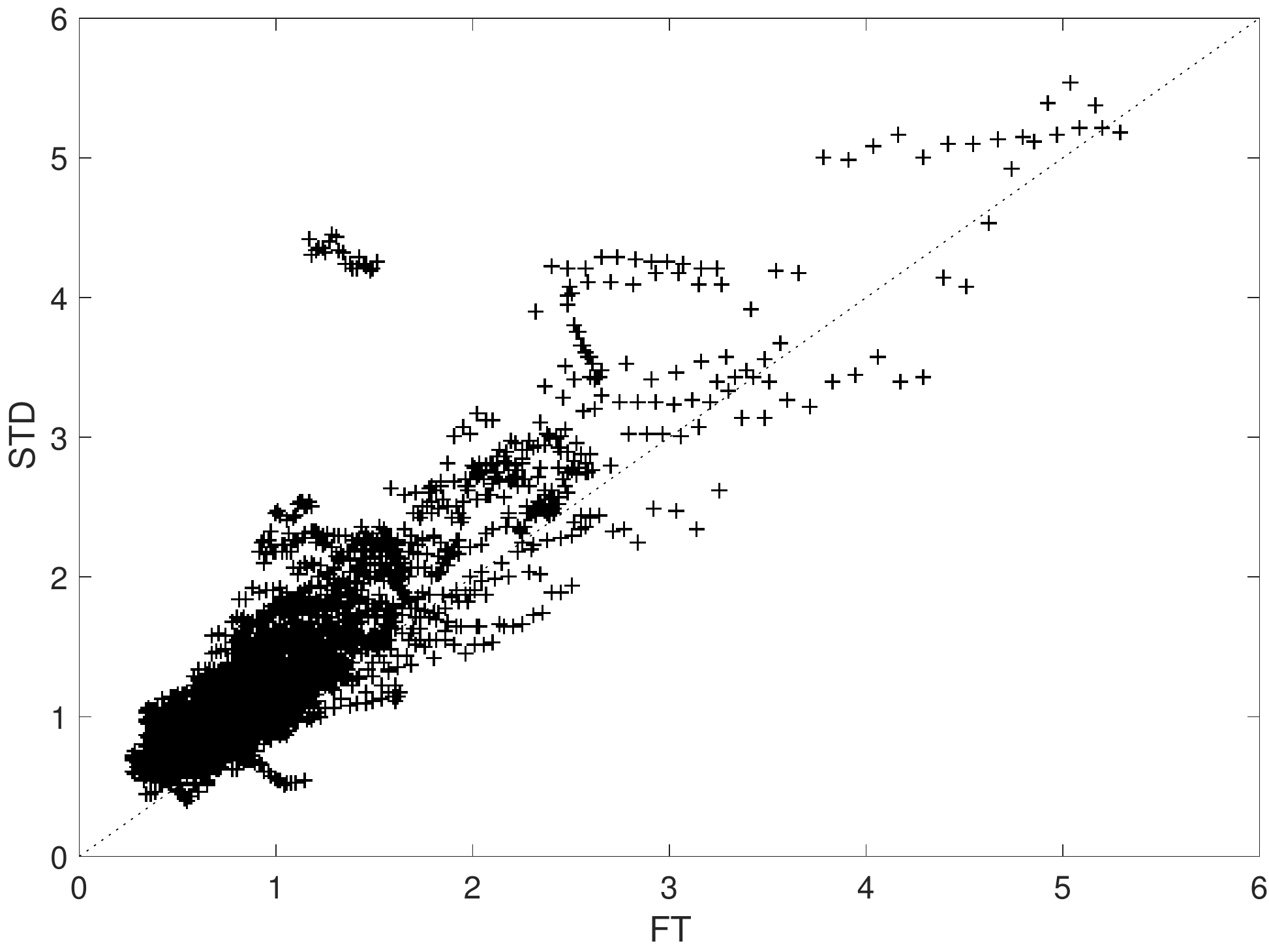}
}
\subfloat[Weekly]{
  \includegraphics[width=0.6\columnwidth]{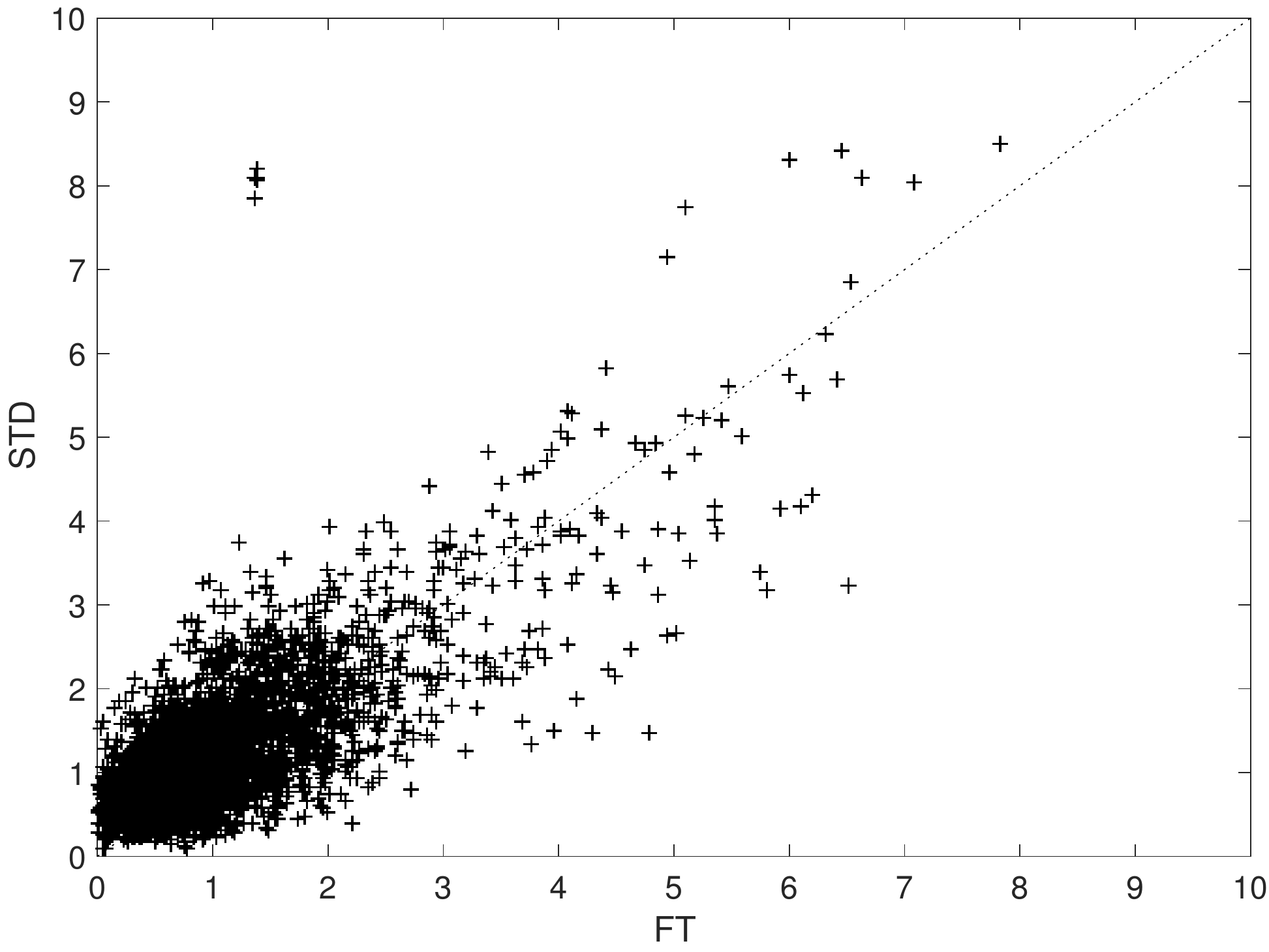}
}
\caption{Scatter plot of values provided by FT and STD volatilities}
\label{fig:scatter}
\end{figure*}

\section{Experimental Results}

\begin{figure}[h!]
\centering
\includegraphics[width=0.8\columnwidth]{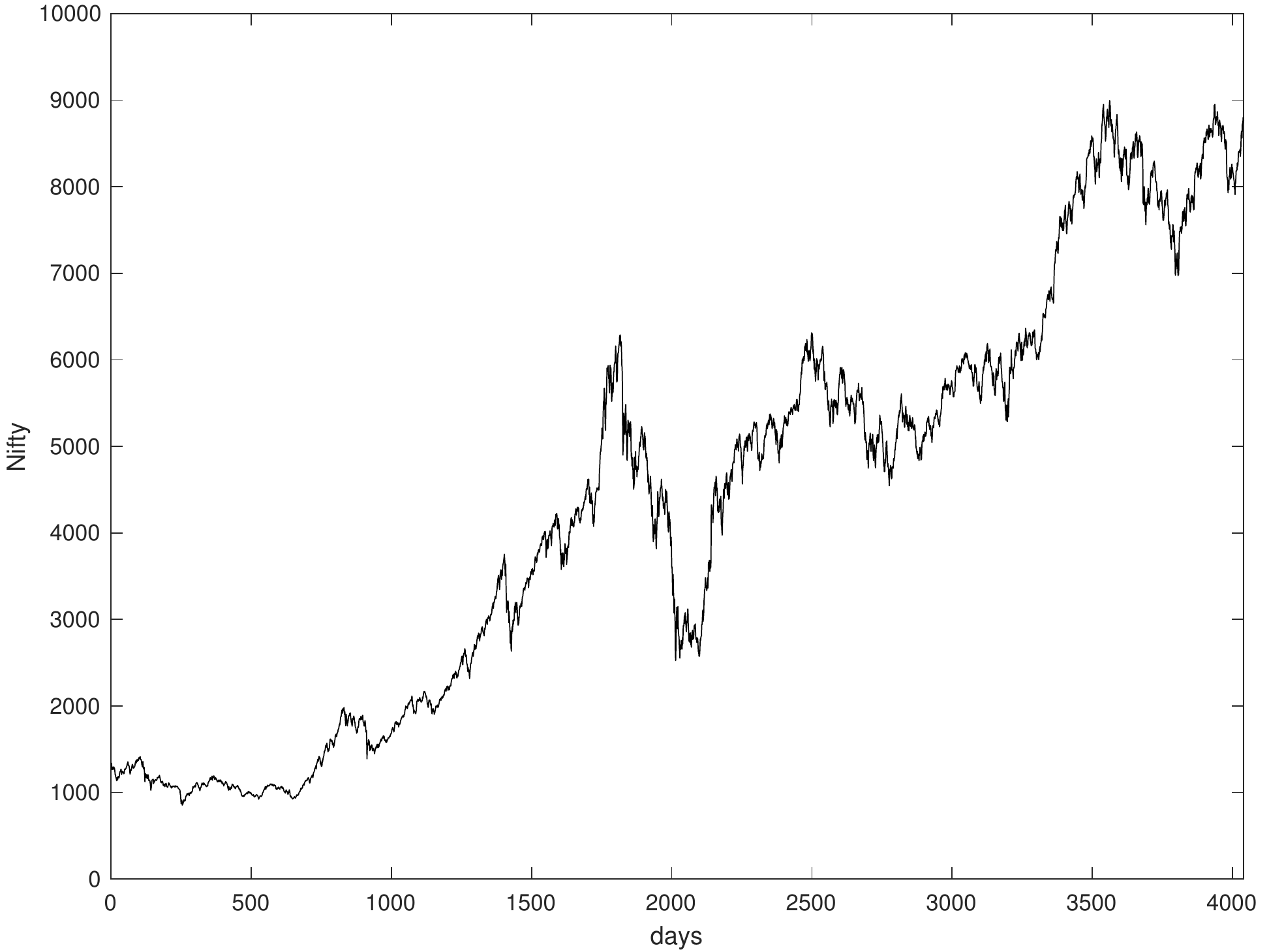}
\caption{The performance of Nifty 50}
\label{fig:nifty}
\end{figure}

\begin{figure}[h!]
\centering
\includegraphics[width=0.8\columnwidth]{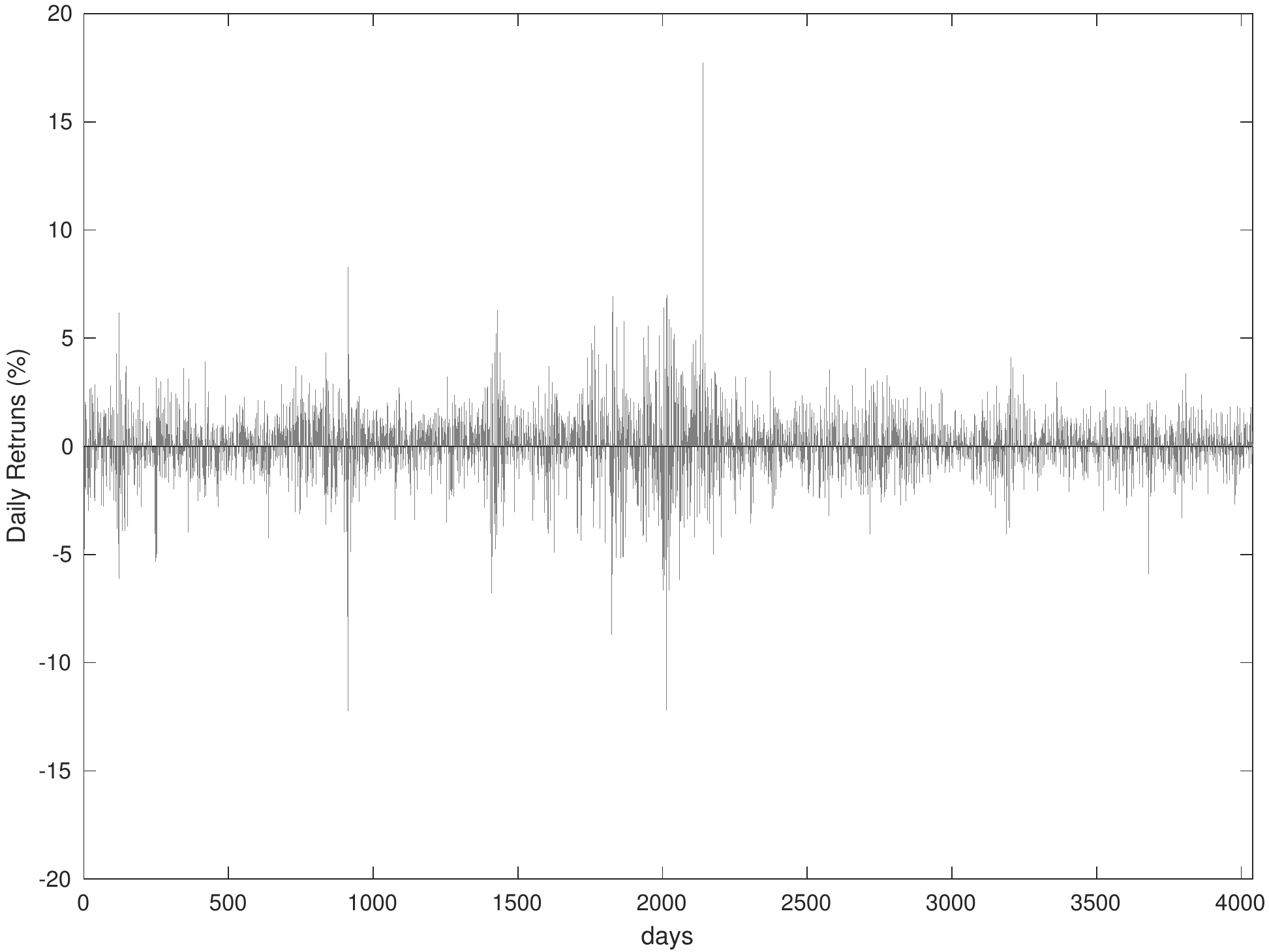}
\caption{Daily returns}
\label{fig:daily}
\end{figure}

For testing the quality of the proposed model, we considered the application to NIFTY 50 that is the benchmark index used by National Stock Exchange of India for the equity market. The period used for our analyses is 20th September 2000 and 9th February 2017, entailing 4040 days. The index performance and daily returns during this period are respectively plotted in Fig.\ref{fig:nifty} and Fig.\ref{fig:daily}.

\begin{figure}[t!]
\centering
\subfloat[Yearly]{
  \includegraphics[width=1\columnwidth]{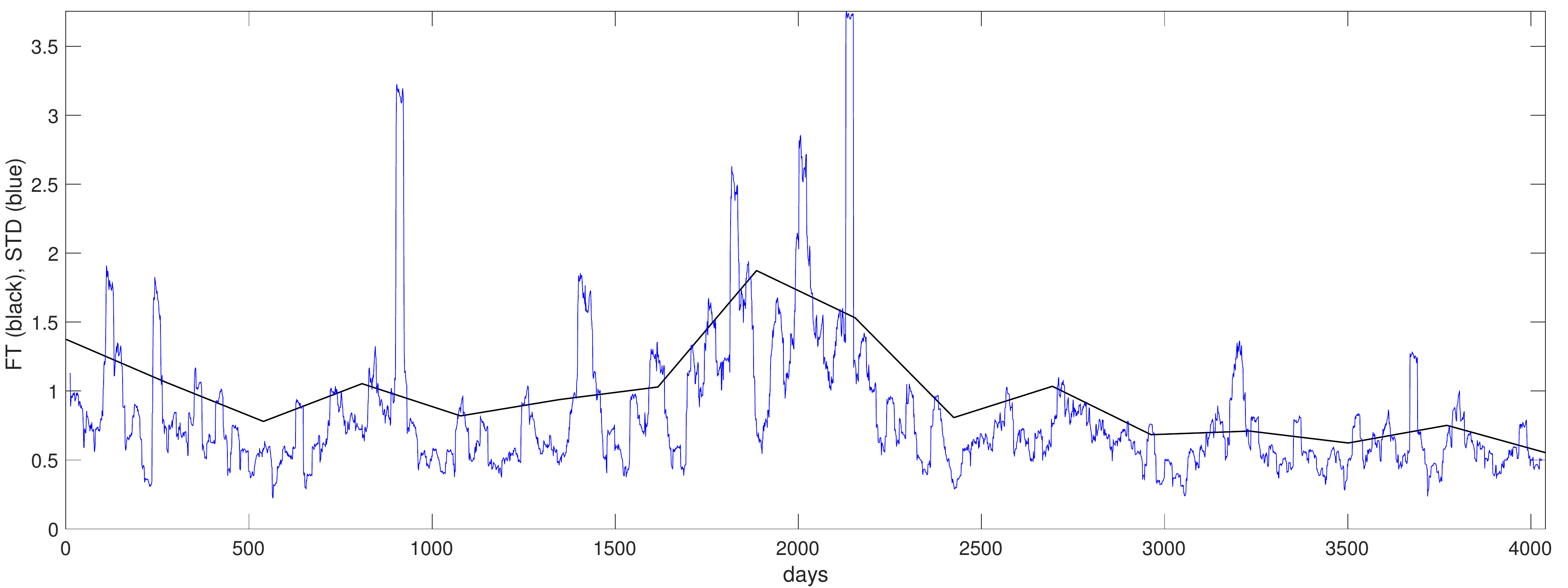}
}
\\
\subfloat[Monthly]{
  \includegraphics[width=1\columnwidth]{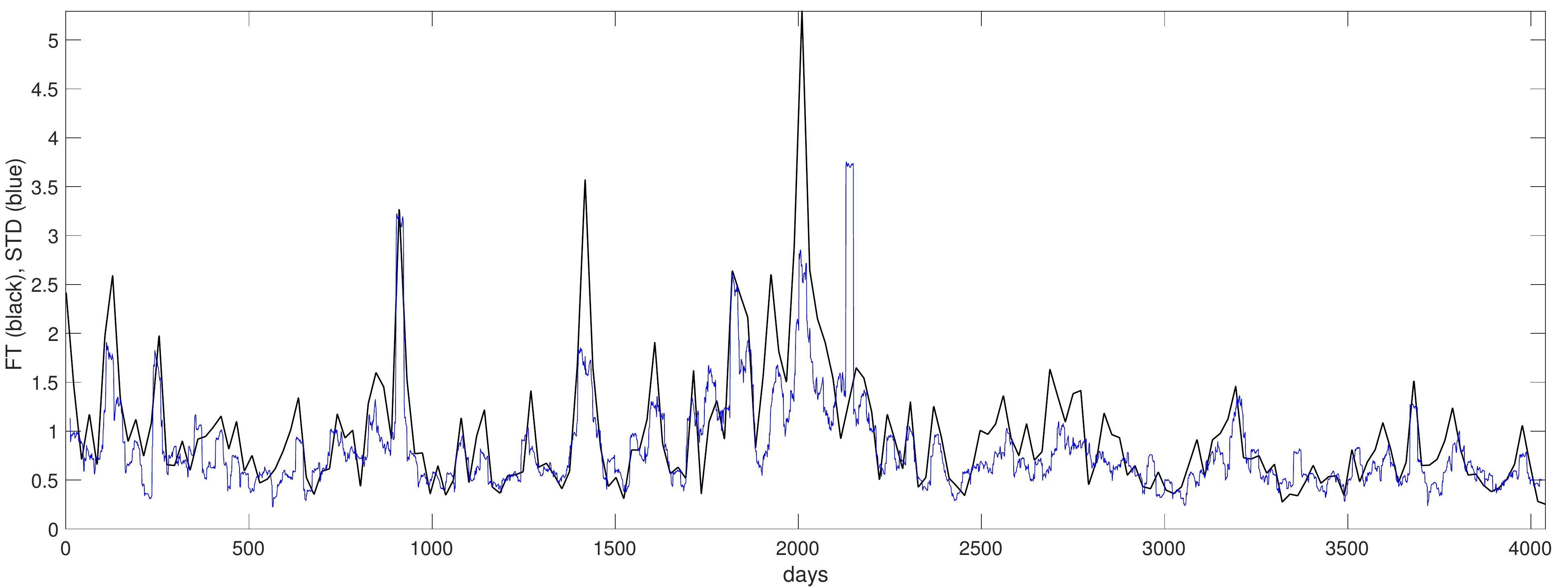}
}
\\
\subfloat[Weekly]{
  \includegraphics[width=1\columnwidth]{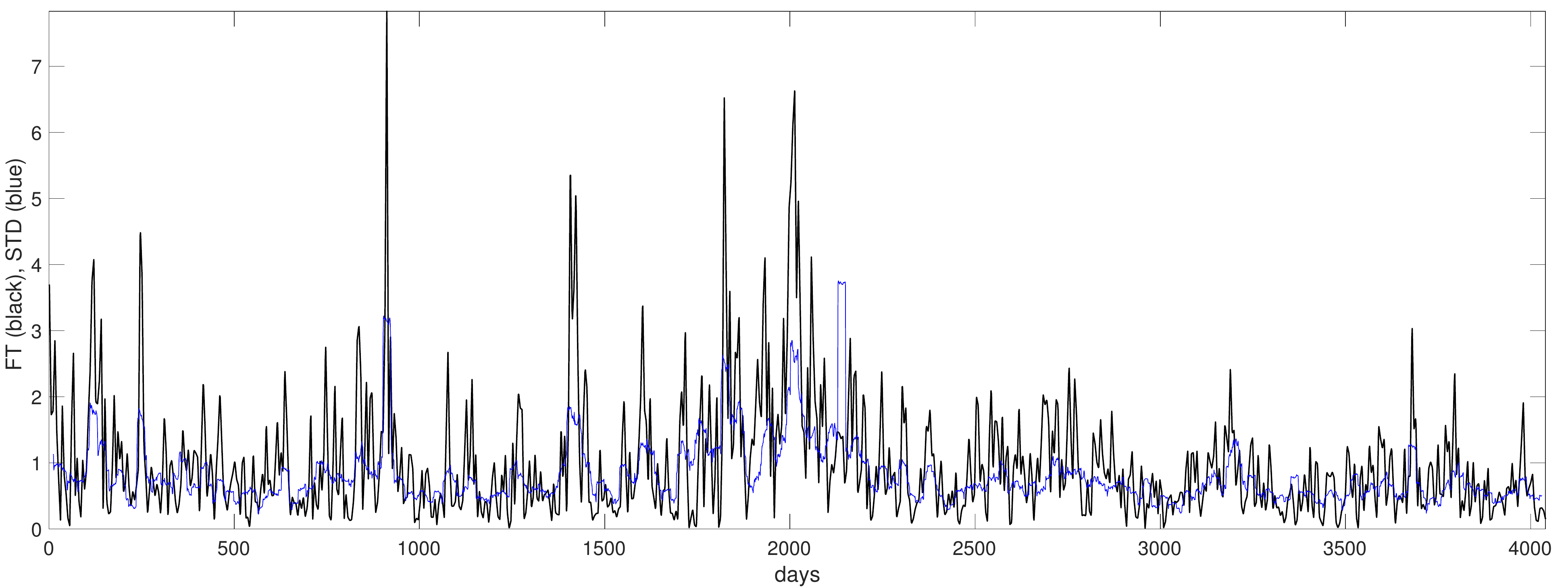}
}
\caption{Point-wise comparison of FT (continuous) and STD (dotted) volatility.}
\label{fig:pointwise}
\end{figure}

Volatility computed by means of fuzzy transform (and its inverse) is shown in Fig.\ref{fig:FT_vol} at different time horizons, i.e. yearly (252 days), monthly (21 days) and weekly (5 days), where on y-axis we report $\rho = r - \beta$, that are  \emph{baseline-adjusted} daily returns, where $\beta$ is the FT-baseline.
Volatility measured by means of fuzzy transform looks smoother if compared to volatility computed as standard deviation and outlined in Fig.\ref{fig:STD_vol}. In this case we reported \emph{mean-adjusted} daily return, that are given as $\rho = r - \mu$. This property is due to the fuzzy partition that offers a gradual transition from one node to the following. Having fixed nodes is also the reason for which FT volatility is better centered than STD volatility. This effect is due to different procedures: FT makes use of fixed nodes over the time, while STD makes use of a rolling windows moved ahead step by step. This is a minor effect that can be easily compensated by aligning the two measures by sliding half of the period one of the two (backward the STD or forward the FT). In order to reduce the effect of delay in comparing the two measures, we used a centered rolling window for the STD volatility. The point-wise comparison is given by Fig.\ref{fig:pointwise}. Although smoother, FT volatility seems to follow the trend given by the series of daily returns. Thus it is related to values given by STD volatility.
This aspect is further investigated by looking at the scatter plots given by Fig.\ref{fig:scatter}. The figures depicts the high correlation shown by both volatility measures, that is 0.9167 on yearly basis, 0.8755 on monthly basis and 0.7523 on weekly basis (Pearson's correlation coefficients). This is interesting if we consider that FT volatility is computed using respectively only 16 (yearly), 192 (monthly) and 808 (weekly) points against the over 4000 points used to outline the STD volatility. Correlation becomes weaker by shortening the time horizon, and FT provides in general a higher level of volatility. It is worth to note that STD volatility measure provides an interpretable result in terms of return probability distribution, although this stands when returns are Gaussian distributed.

\section{Conclusions}

In this paper we proposed an alternative way to measure volatility that is based on fuzzy transform and its inverse. Preliminary experimentation on NIFTY 50 stock index shows the measure obtained offers a more regular/stable and better centered measure of volatility when compared to volatility measured by standard deviation. The nature of FT value is instead related to difference between absolute and signed returns. Both are able to mitigate the effect of extreme returns. This work is preliminary and future research should further investigate analytical and statistical properties of this measure.




%
\bibliographystyle{myIEEETran}
\bibliography{references.bib}

\begin{thebibliography}{1}
\providecommand{\url}[1]{#1}
\csname url@samestyle\endcsname
\providecommand{\newblock}{\relax}
\providecommand{\bibinfo}[2]{#2}
\providecommand{\BIBentrySTDinterwordspacing}{\spaceskip=0pt\relax}
\providecommand{\BIBentryALTinterwordstretchfactor}{4}
\providecommand{\BIBentryALTinterwordspacing}{\spaceskip=\fontdimen2\font plus
\BIBentryALTinterwordstretchfactor\fontdimen3\font minus
  \fontdimen4\font\relax}
\providecommand{\BIBforeignlanguage}[2]{{%
\expandafter\ifx\csname l@#1\endcsname\relax
\typeout{** WARNING: IEEEtran.bst: No hyphenation pattern has been}%
\typeout{** loaded for the language `#1'. Using the pattern for}%
\typeout{** the default language instead.}%
\else
\language=\csname l@#1\endcsname
\fi
#2}}
\providecommand{\BIBdecl}{\relax}
\BIBdecl

\bibitem{JAE:JAE687}
C.~W.~J. Granger, ``Some comments on risk,'' \emph{Journal of Applied
  Econometrics}, vol.~17, no.~5.

\bibitem{Luce1981}
R.~D. Luce, ``Correction to `several possible measures of risk','' \emph{Theory
  and Decision}, vol.~13, no.~4, pp. 381--381, 1981. doi: 10.1007/BF00126971

\bibitem{Luce1980}
R.~D. Luce, ``Several possible measures of risk,'' \emph{Theory and Decision},
  vol.~12, no.~3, pp. 217--228, 1980. doi: 10.1007/BF00135033

\bibitem{Perfilieva:2006}
I.~Perfilieva, ``Fuzzy transforms: Theory and applications,'' \emph{Fuzzy Sets
  Syst.}, vol. 157, no.~8, pp. 993--1023, Apr. 2006. doi:
  10.1016/j.fss.2005.11.012

\bibitem{Loia2017209}
V.~Loia, S.~Tomasiello, and A.~Vaccaro, ``Using fuzzy transform in multi-agent
  based monitoring of smart grids,'' \emph{Information Sciences}, vol. 388-389,
  pp. 209--224, 2017. doi: 10.1016/j.ins.2017.01.022

\bibitem{Tomasiello2015}
M.~Gaeta, V.~Loia, and S.~Tomasiello, ``Multisignal 1-d compression by
  f-transform for wireless sensor networks applications,'' \emph{Appl. Soft
  Comput.}, vol.~30, no.~C, pp. 329--340, May 2015. doi:
  10.1016/j.asoc.2014.11.061

\bibitem{Troiano201211}
L.~Troiano and P.~Kriplani, ``A mean-reverting strategy based on fuzzy
  transform residuals,'' 2012. doi: 10.1109/CIFEr.2012.6327766 pp. 11--17.

\bibitem{Troiano2011121}
L.~Troiano and P.~Kriplani, ``Supporting trading strategies by inverse fuzzy
  transform,'' \emph{Fuzzy Sets and Systems}, vol. 180, no.~1, pp. 121--145,
  2011. doi: 10.1016/j.fss.2011.05.004

\bibitem{Troiano2010379}
L.~Troiano, ``Fuzzy co-transform and its application to time series,'' 2010.
  doi: 10.1109/SOCPAR.2010.5686735 pp. 379--384.

\end{thebibliography}

\end{document}